# The relaxation of thermal and high-frequency transverse phonons in the semiconductor cubic crystals

I G Kuleyev, I I Kuleyev, A P Tankeyev and I Yu Arapova

Institute of Metal Physics, Ural Branch of the Russian Academy of Sciences, Ekaterinburg 620219, Russian Federation

## 1. Introduction

Investigations of the phonons relaxation mechanisms in the anharmonic scattering processes is of interest not only for studying the lattice thermal conductivity of crystals [1], but also such kinetic effects, as the ultrasound absorption [2] and the electron-phonon drag thermoelectric power [3]. Experimental investigations of the isotope effects in the thermal conductivity of germanium, silicon and diamond [4-8], and the theoretical analysis of these results [9-10] showed, that the normal processes (N-processes) of phonon-phonon scattering play an important role in the lattice thermal conductivity of isotopically enriched and chemically pure crystals. These processes in which the full momentum of the colliding phonons is conserved, together with the boundary scattering of phonons are the main mechanisms limiting the maximum values of the thermal conductivity in highly isotopically enriched crystals [9,10].

Calculations of the thermal conductivity within the framework of the relaxation method [4–12] usually employ expressions for the phonon relaxation rates in the $N$ processes obtained in the long-wavelength approximation, $z_{q\lambda} = \hbar\omega_{q\lambda}/k_B T \ll 1$ (where $\hbar\omega_{q\lambda}$ is the energy of a phonon with the wave vector **q** and the polarization λ). This approach is well justified for the calculation of the long-wave ultrasound absorption coefficient, and also the phonon-drag component thermo power since in semiconductor crystals electrons can interact only with the long-wave phonons. However, the main contribution to the lattice thermal conductivity in crystals with the natural isotope composition is due to the thermal phonons with $z_{q\lambda} \approx 1$, and in isotopically enriched crystals—due to the phonons with $z_{q\lambda} \approx 2-4$. Therefore the long-wave approximation for the phonon relaxation rate is not correct for a calculation of the thermal conductivity. In the framework of this approach, quantities determining the intensity of anharmonic scattering processes are the fitting parameters of the theory; they are determined from a comparison of the theoretical calculation and experimental data [4-12]. For the estimation of the probability of anharmonic scattering processes the model of isotropic medium was usually used. This model is inadequate for germanium, silicon and diamond crystals, as well as for such popular objects of research as InSb, GaAs, CaF$_2$, etc. possessing cubic symmetry with essential anisotropy of elastic moduli, both of the second and the third orders.

The aim of our study is to calculate the transverse phonon relaxation rates in the entire range of wave vectors in the anharmonic scattering processes using the experimentally obtained values of the elastic moduli of the second and the third order. The determination of the temperature and wave-vector dependences of the phonon relaxation rates in cubic crystals enables us, first, to obtain the attenuation coefficients both for the long-wave ($z_{q\lambda} << 1$) and the short-wave ($z_{q\lambda} >> 1$) ultrasound, second, to select the effective mechanisms of the thermal phonon relaxation and avoid the arbitrariness related to the choice of two fitting parameters of the theory for the calculation of the thermal conductivity of these crystals.

In this paper we present the results of our study of the transverse phonons relaxation for Landau-Rumer mechanism in isotropic medium, as well as in Ge, Si, diamond, InSb, GaSb, GaAs, LiF, CaF$_2$ and SrF$_2$ crystals possessing a cubic symmetry are presented. The relations between the elastic moduli which are required for a transition to the isotropic medium model are considered. The relaxation rates of the transverse thermal and high-frequency phonons for the above-mentioned cubic crystals are calculated. The calculations allow us to determine the values of the transverse phonon relaxation rates in anharmonic scattering processes as well as the ultrasonic attenuation coefficients from the experimental values of the elastic moduli. The obtained results are compared to those derived previously in the model of isotropic medium. We also consider the long-wavelength limit, the transition to the model of isotropic medium, and the dependences of the phonon relaxation rate on the temperature and the wave vector for cubic crystals. It is shown that the dependence of the transverse phonon relaxation rate on the wave vector can be determined using the temperature dependence of the high-frequency ultrasonic attenuation coefficient.

**2. Elastic energy of the cubic symmetry crystals**

The expression for the elastic energy of the cubic symmetry crystals up to the third order on the deformation tensor components $\eta_{ij}$ was received in [13]. Let us transform it analogously to the Ref. [14], where it was make for the isotropic medium model. For this purpose we shall express components of the deformation tensor $\eta_{ik}$ via the distortion tensor components $\xi_{ik}$ [15]:

$$\eta_{ik} = \frac{1}{2}\left( \frac{\partial u_i}{\partial x_k} + \frac{\partial u_k}{\partial x_i} + \sum_j \frac{\partial u_j}{\partial x_i}\frac{\partial u_j}{\partial x_k} \right) \equiv \frac{1}{2}\left( \xi_{ik} + \xi_{ki} + \sum_j \xi_{ji}\xi_{jk} \right). \quad (1)$$

Let us present the value of $\xi_{ik}$ in the standard form [14]:

$$\xi_{ik} = \frac{\partial u_i}{\partial x_k} = i\sum_q \left( \frac{\hbar}{2\rho V \omega_{q\lambda}} \right)^{1/2} e_i^\lambda \mathbf{q}_k \exp(i\mathbf{qr})\left( b_{q\lambda} - b_{-q\lambda}^+ \right), \quad (2)$$

where $b_{-q\lambda}^+$ and $b_{q\lambda}$ are the operators of production and annihilation of phonons, $\rho$ . is the density, $V$ is the normalizing volume, and $\mathbf{e}$ is the polarization vector, $\omega_{q\lambda}$ is the frequency of the phonon with the wave vector $q$ and the polarization $\lambda$. In terms of the operators $b^+_{-q\lambda}$, $b_{q\lambda}$ an expression for the elastic energy determining the probabilities of various anharmonic three-phonon scattering processes via the elastic moduli of the second and the third order can be expressed as

$$W = \frac{1}{6} \sum_{\substack{\mathbf{q}_1, \mathbf{q}_2, \mathbf{q}_3 \\ \lambda_1 \lambda_2 \lambda_3}} F^{\lambda_1 \lambda_2 \lambda_3}_{\mathbf{q}_1 \mathbf{q}_2 \mathbf{q}_3} (b_{q_1 \lambda_1} - b^+_{-q_1 \lambda_1})(b_{q_2 \lambda_2} - b^+_{-q_2 \lambda_2})(b_{q_3 \lambda_3} - b^+_{-q_3 \lambda_3}),$$ (3)

where $F^{\lambda_1 \lambda_2 \lambda_3}_{\mathbf{q}_1 \mathbf{q}_2 \mathbf{q}_3} = -i \left(\frac{\hbar}{2\rho V}\right)^{3/2} (\omega_{q_1 \lambda_1} \omega_{q_2 \lambda_2} \omega_{q_3 \lambda_3})^{-1/2} V^{\lambda_1 \lambda_2 \lambda_3}_{\mathbf{q}_1 \mathbf{q}_2 \mathbf{q}_3} \delta(\mathbf{q}_1 + \mathbf{q}_2 + \mathbf{q}_3, 0),$ (4)

$$V^{\lambda_1\lambda_2\lambda_3}_{\mathbf{q}_1\mathbf{q}_2\mathbf{q}_3} = \widetilde{c}_{111} \sum_i e_{1i} q_{1i} e_{2i} q_{2i} e_{3i} q_{3i} + c_{123}(\mathbf{e}_1\mathbf{q}_1)(\mathbf{e}_2\mathbf{q}_2)(\mathbf{e}_3\mathbf{q}_3) +$$
$$+ \widetilde{c}_{112} \sum_i [(\mathbf{e}_1\mathbf{q}_1) e_{2i} q_{2i} e_{3i} q_{3i} + (\mathbf{e}_2\mathbf{q}_2) e_{1i} q_{1i} e_{3i} q_{3i} + (\mathbf{e}_3\mathbf{q}_3) e_{1i} q_{1i} e_{2i} q_{2i}] +$$
$$+ c_{144}[(\mathbf{e}_1\mathbf{q}_1)(\mathbf{e}_2\mathbf{q}_3)(\mathbf{e}_3\mathbf{q}_2) + (\mathbf{e}_2\mathbf{q}_2)(\mathbf{e}_1\mathbf{q}_3)(\mathbf{e}_3\mathbf{q}_1) + (\mathbf{e}_3\mathbf{q}_3)(\mathbf{e}_1\mathbf{q}_2)(\mathbf{e}_2\mathbf{q}_1)] +$$
$$+ (c_{12} + c_{144})[(\mathbf{e}_1\mathbf{q}_1)(\mathbf{e}_2\mathbf{e}_3)(\mathbf{q}_3\mathbf{q}_2) + (\mathbf{e}_2\mathbf{q}_2)(\mathbf{e}_1\mathbf{e}_3)(\mathbf{q}_3\mathbf{q}_1) + (\mathbf{e}_3\mathbf{q}_3)(\mathbf{e}_1\mathbf{e}_2)(\mathbf{q}_1\mathbf{q}_2)] +$$
$$+ c_{456}[(\mathbf{e}_1\mathbf{q}_3)(\mathbf{e}_2\mathbf{q}_1)(\mathbf{e}_3\mathbf{q}_2) + (\mathbf{e}_1\mathbf{q}_2)(\mathbf{e}_2\mathbf{q}_3)(\mathbf{e}_3\mathbf{q}_1)] +$$
$$+ (c_{44} + c_{456})[(\mathbf{e}_1\mathbf{q}_2)(\mathbf{q}_1\mathbf{q}_3)(\mathbf{e}_2\mathbf{e}_3) + (\mathbf{e}_2\mathbf{q}_3)(\mathbf{q}_1\mathbf{q}_2)(\mathbf{e}_1\mathbf{e}_3) + (\mathbf{e}_3\mathbf{q}_1)(\mathbf{q}_2\mathbf{q}_3)(\mathbf{e}_1\mathbf{e}_2) +$$
$$+ (\mathbf{e}_1\mathbf{q}_3)(\mathbf{q}_1\mathbf{q}_2)(\mathbf{e}_2\mathbf{e}_3) + (\mathbf{e}_2\mathbf{q}_1)(\mathbf{q}_2\mathbf{q}_3)(\mathbf{e}_1\mathbf{e}_3) + (\mathbf{e}_3\mathbf{q}_2)(\mathbf{q}_1\mathbf{q}_3)(\mathbf{e}_1\mathbf{e}_2)] +$$
$$+ \widetilde{c}_{155} \sum_i \{e_{1i} e_{2i} e_{3i} (q_{1i}(\mathbf{q}_2\mathbf{q}_3) + q_{2i}(\mathbf{q}_1\mathbf{q}_3) + q_{3i}(\mathbf{q}_1\mathbf{q}_2)) +$$
$$+ e_{1i} q_{1i}[e_{2i} q_{3i}(\mathbf{e}_3\mathbf{q}_2) + e_{3i} q_{2i}(\mathbf{e}_2\mathbf{q}_3)] + e_{2i} q_{2i}[e_{1i} q_{3i}(\mathbf{e}_3\mathbf{q}_1) + e_{3i} q_{1i}(\mathbf{e}_1\mathbf{q}_3)] +$$
$$+ e_{3i} q_{3i}[e_{1i} q_{2i}(\mathbf{e}_2\mathbf{q}_1) + e_{2i} q_{1i}(\mathbf{e}_1\mathbf{q}_2)] \} + [\widetilde{c}_{155} + \Delta C] \sum_i q_{1i} q_{2i} q_{3i} [e_{1i}(\mathbf{e}_2\mathbf{e}_3) +$$
$$+ e_{2i}(\mathbf{e}_1\mathbf{e}_3) + e_{3i}(\mathbf{e}_1\mathbf{e}_2)].$$

In expression (4) the contributions containing the elastic moduli:

$\widetilde{c}_{111} = c_{111} - 3c_{112} + 2c_{123} + 12c_{144} - 12c_{155} + 16c_{456}$, $\Delta C = c_{11} - c_{12} - 2c_{44}$

$\widetilde{c}_{112} = c_{112} - c_{123} - 2c_{144},$  $\widetilde{c}_{155} = c_{155} - c_{144} - 2c_{456}$ (5)

are typical of cubic crystals and distinguish them from the isotropic medium. The values of the thermodynamic moduli of the third order $c_{ijk}$ are experimentally determined in Ref. [16-19]. Therefore in expression (4) for the elastic moduli of the third order $c_{ijk}$ we have used normalization accepted by Bragger [20]. The passage to the Birch normalization $c^B_{ijk}$ [13] is provided by the substitution

$c_{111} = 6c^B_{111}$, $c_{112} = 2c^B_{112}$, $c_{123} = c^B_{123}$, $c_{144} = \frac{1}{2}c^B_{144}$, $c_{155} = \frac{1}{2}c^B_{155}$, $c_{456} = \frac{1}{4}c^B_{456}$. (6)

The passage to the Taker and Rempton normalization of the elastic moduli [14] is realized by the substitution:

$$c_{ijk}^{TR} = \frac{1}{6} c_{ijk}. \qquad (6a)$$

Let's notice, that in [19] for the thermodynamic elastic modulus $c_{456}$ the transition factor to the Birch normalization is incorrectly specified: the factor 1/8 was borrowed instead the factor 1/4 (see also [20], [21]). This transition factor was used in Ref. [22] for calculations of the phonon relaxation rate in the Landau-Rumer mechanism for Ge, Si and diamond crystal. For the correction of this inaccuracy the values of modulus $c_{456}$ (see Table I, Ref. [22]) in Taker and Rempton normalization [14], it is necessary to multiply by a factor two. This correction did not lead to qualitative change of the results, obtained in Ref. [22] for the relaxation rates for Ge, Si and diamond crystal (compare Figs. 3a, 3b and 5 in this paper with Fig. 4 of Ref. [22]). The relations between the isotropic and anisotropic scattering contributions of the relaxation rates above-mentioned crystals will change quantitatively only. Therefore below we reproduce general calculations, which leads to the expressions for the relaxation rates in cubic crystals in term of the thermodynamic modulus $c_{ijk}$ and verify results obtained in Ref. [22] for Ge, Si and diamond.

Let us consider the transition from the elastic energy of the cubic symmetry crystals to the elastic energy of the isotropic medium $W_{iso}$. For this purpose we shall compare the expression (4) with formula (4.22) of Ref. [14]. From the condition $\Delta W = W_k - W_{iso} = 0$ we shall find:

$\Delta C = c_{11} - c_{12} - 2c_{44} = 0$,

$\tilde{c}_{155} = c_{155} - c_{144} - 2 c_{456} = 0$, \qquad (7)

$\tilde{c}_{112} = c_{112} - c_{123} - 2 c_{144} = 0$,

$\tilde{c}_{111} = c_{111} - 3 c_{112} + 2 c_{123} + 12 c_{144} - 12 c_{155} + 16 c_{456} = 0$.

Below we analyze how the conditions (7) are satisfied in the case of various cubic crystals, using the results of measurements of the second and the third order elastic modules reported in [16-20] and summarized in Table I.

**Table I.** Elastic moduli of the first and second order. Thermodynamic quantities $c_{ijk}$ are presented according to [17].

|  | $\rho$, g/cm$^3$ | $c_{11}$ | $c_{12}$ | $c_{44}$ | $c_{111}$ | $c_{112}$ | $c_{123}$ | $c_{144}$ | $c_{155}$ | $c_{456}$ |
|---|---|---|---|---|---|---|---|---|---|---|
| Ge | 5,32 | 1,289 | 0,483 | 0,671 | -7,10 | -3,89 | -0,18 | -0,23 | -2,92 | -0,53 |
| Si | 2,33 | 1,657 | 0,638 | 0,796 | -8,25 | -4,51 | -0,64 | 0,12 | -3,10 | -0,64 |
| Diamond | 3,51 | 10,76 | 1,25 | 5,758 | -62,6 | -22,6 | 1,12 | -6,74 | -28,6 | -8,23 |
| InSb | 5,76 | 0.672 | 0.367 | 0.302 | -3.71 | -2.83 | -1.15 | 0.21 | -1.41 | 0.03 |
| GaSb | 5.62 | 0.885 | 0.404 | 0.433 | -4.75 | -3.08 | -0.44 | 0.5 | -2.16 | -0.25 |
| GaAs | 5,31 | 0.192 | 0.0599 | 0.0538 | -6.22 | -3.87 | -0.57 | 0.02 | -2.69 | -0.39 |
| CaF$_2$ | 3,18 | 1.644 | 0.502 | 0.347 | -12.46 | -4.0 | -2.54 | -1.24 | -2.14 | -0.748 |
| LiF | 2,6 | 1.144 | 0.426 | 0.628 | -1.92 | -0.33 | -0.04 | 0.1 | -0.33 | 0.043 |
| SrF$_2$ | 2.44 | 1.24 | 0.43 | 0.31 | -8.21 | -3.09 | -1.81 | -0.95 | -1.75 | -0.42 |

Our analysis showed (see Table 2) that neither the second-order nor the third-order elastic modules for Ge, Si, diamond, InSb, GaSb, GaAs, CaF$_2$, SrF$_2$ crystals satisfy relations (7). For example, the value of ΔC for Ge, Si, InSb and GaSb crystals is negative and its absolute value is of the order of (or more than) the elastic modulus c$_{12}$, and the value of $\widetilde{c}_{155}$ is in the range of 50 % to 100 % of the value c$_{155}$. For CaF$_2$ and SrF$_2$ crystals the elastic modulus $\widetilde{c}_{155}$ has the sign opposite to that of c$_{155}$, and surpasses it in the absolute value.

**Table II.** Elastic moduli ΔC, $\widetilde{c}_{111}$, $\widetilde{c}_{112}$, $\widetilde{c}_{155}$, A$_{cub}$ (in units of $10^{12}$ dyn/cm$^2$)

|  | Ge | Si | Diamond | InSb | GaSb | GaAs | CaF$_2$ | LiF | SrF$_2$ |
|---|---|---|---|---|---|---|---|---|---|
| ΔC | -0,54 | -0,57 | -2,01 | -0,29 | -0,385 | 0,025 | 0,45 | -0,54 | 0,185 |
| $\widetilde{c}_{112}$ | -3,25 | -4,1 | -10,24 | -2,09 | -3,64 | -3,34 | 1,019 | -0,49 | 0,62 |
| $\widetilde{c}_{111}$ | 28,01 | 32,4 | 138,1 | 18,19 | 31,5 | 30,5 | -6,71 | 4,84 | 0,32 |
| $\widetilde{c}_{155}$ | -1,63 | -1,9 | -5,4 | -1,29 | -2,16 | -1,9 | 0,59 | -0,52 | 0,039 |
| A$_{cub}$ | -0,08 | 0,71 | -27,9 | 1,57 | 1,7 | -1,29 | -3,93 | 2,68 | -2,22 |

The maximum discrepancy with the model of isotropic medium is observed for the elastic modulus $\widetilde{c}_{111}$: this value not only significantly exceeds the other third-order elastic moduli of germanium, silicon, and diamond, but it has the sign opposite to that of the $c_{111}$ value. The similar situation takes place for InSb, GaSb, GaAs and LiF crystals. On the contrary, for CaF$_2$ crystal the value and sign of elastic modulus $\widetilde{c}_{111}$ and c$_{111}$ coincides. As the elastic moduli $\widetilde{c}_{111}$, $\widetilde{c}_{112}$ and $\widetilde{c}_{155}$ distinguish the cubic crystals from the isotropic medium then the greatest deviation from isotropic medium model takes place for those relaxation rates in which the members with the elastic module $\widetilde{c}_{111}$ make the contribution. The expressions (3) and (4) allow to investigate various mechanisms of the three-phonon scattering processes in the cubic crystals. The squared modulus of the matrix element $V_{q_1 q_2 q_3}^{\lambda_1 \lambda_2 \lambda_3}$ determines the probabilities of the anharmonic scattering processes of phonons. Further it is used for calculating of the phonons relaxation rates in the normal scattering processes.

### 3. Mechanism Landau-Rumer for transverse phonons in cubic crystals

According to the Landau–Rumer mechanism [23] the relaxation of long-wavelength transverse phonons ($\hbar\omega_{qT} \ll k_B T$) consists of transverse and longitudinal phonons merging with the formation of a new longitudinal phonon. An analysis of this relaxation mechanism occurring at a sufficiently low temperature yields an expression for the relaxation rate [23]

$$\nu_{phN}^{LR} = B_{T0} z_1 T^5, \qquad z_1 = \frac{\hbar \omega_{q_1}^T}{k_B T}, \tag{8}$$

where $B_{T0}$ is a coefficient depending on the second- and third-order elastic moduli, atomic masses, and lattice parameters. In the theory of lattice thermal conductivity, this coefficient is considered as the fitting parameter determined from a comparison of theoretical calculation and experimental data [4-12]. In Ref. [12,15,24], an analysis of the phonon relaxation according to the Landau–Rumer (LR) mechanism was restricted to the long-wavelength approximation ($q_1 \ll q_2, q_3$ ($z_1 \ll 1$)) and the model of isotropic medium. The same approximations were used for the relaxation rate of thermal phonons in the calculations of phonon thermal conductivity, although the condition $z_1 \ll 1$ is not valid for thermal phonons. Obviously, the previous results [4-12] have to be refined to reject the long-wavelength approximation and the model of isotropic medium.

Below we shall adduce calculation of the relaxation rate of $\nu_{phN}^{TLL}$ for cubic crystals which will allow to determine the coefficient $B_{T0}$ using the known values of the second- and third-order elastic moduli and refine the dependence of the relaxation rate $\nu_{phN}^{TLL}$ on the temperature and wave vector of thermal phonons ($z_1 > 1$). According to Ref. [12], an expression for the phonon relaxation rate can be written as

$$\nu_{phN}(q_1,\lambda_1) = \frac{\pi \hbar^4}{(2\rho k_B T)^3} \frac{1}{V} \sum_{\substack{\mathbf{q}_2 \mathbf{q}_3 \\ \lambda_2 \lambda_3}} \frac{sh\left(\frac{z_1}{2}\right) \cdot \delta_{\mathbf{q}_1+\mathbf{q}_2+\mathbf{q}_3,0}}{z_1 z_2 z_3 sh\left(\frac{z_2}{2}\right) sh\left(\frac{z_3}{2}\right)} \left|V_{\mathbf{q}_1\mathbf{q}_2\mathbf{q}_3}^{\lambda_1\lambda_2\lambda_3}\right|^2 \quad (9)$$

$$\left[2\delta\left(\omega_{q_1\lambda_1} + \omega_{q_2\lambda_2} - \omega_{q_3\lambda_3}\right) + \delta\left(\omega_{q_1\lambda_1} - \omega_{q_2\lambda_2} - \omega_{q_3\lambda_3}\right)\right]$$

The transverse phonons can be involved only in the process of "merging" (reflected by the first term in square brackets), while the probability of decay (second term in square braces) is zero because otherwise the law of energy conservation cannot be satisfied.

The process of phonon merging obeys the following relations:

$(\mathbf{e}_1\mathbf{q}_1)=0$, $(\mathbf{e}_2\mathbf{q}_2)=q_2$, $(\mathbf{e}_3\mathbf{q}_3)=q_3$, $(\mathbf{e}_1\mathbf{q}_3)=(\mathbf{e}_1\mathbf{q}_2)$, $\mathbf{q}_3=\mathbf{q}_2+\mathbf{q}_1$. (10)

Using expression (4) and taking into account relations (10), one can readily obtain a formula for the matrix element according to the Landau–Rumer mechanism:

$$V_{\mathbf{q}_1\mathbf{q}_2\mathbf{q}_3}^{TLL} = A\frac{(\mathbf{e}_1\mathbf{q}_2)}{q_2 q_3}\left[q_2^2 + (\mathbf{q}_1\mathbf{q}_2)\right]\left[q_1^2 + 2(\mathbf{q}_1\mathbf{q}_2)\right] +$$

$$+ \frac{1}{q_2 q_3}\sum_i \{2\tilde{c}_{155} e_{1i} q_{2i} q_{3i} [q_{1i}(\mathbf{q}_2\mathbf{q}_3) + q_{2i}(\mathbf{q}_1\mathbf{q}_3) + q_{3i}(\mathbf{q}_1\mathbf{q}_2)] + \quad (11)$$

$$+ (2\tilde{c}_{155} + \Delta C)[(\mathbf{e}_1\mathbf{q}_2)q_{1i}q_{2i}q_{3i}(q_{1i} + q_{2i}) + e_{1i}q_{1i}q_{2i}q_{3i}(\mathbf{q}_2\mathbf{q}_3)]\} .$$

The first term in this expression corresponds to the isotropic scattering of phonons and the second, to their anisotropic scattering (this term vanishes passing to the model of isotropic medium). For a cubic crystal

$A=A_{cub}=c_{12}+3c_{44}+2c_{144}+4c_{456}$ (12)

Using relations (7) and passing to the isotropic case, we obtain the result of Tucker and Rampton [14]:

$$A = A_{iso} = 1.5\,c_{11} - 0.5\,c_{12} + 2c_{155} = 1.5\,c_{11} - 0.5\,c_{12} + 12\,c_{155}^{TR}. \tag{12a}$$

In the general case, the relaxation rate $v_{phN}^{TLL}$ depends on the direction of propagation of the transverse phonon relative to the crystallographic axes. Below, we simplify the problem and restrict the consideration to one of the symmetric directions ([100], [001], [111], etc.). The *z-axis* and the phonon wave vector $\mathbf{q}_1$ are also oriented along this direction, so that $\mathbf{e}_1$ vector occurs in the *xy* plane (for certainty, coincides with the *x* axis). In this case, the condition $(\mathbf{e}_1\mathbf{q}_1) = 0$ can be supplemented by the conditions $e_{1i}q_{1i}=0$ and the matrix element (11) can be written in terms of the angular variables $\theta_2$ and $\varphi_2$ vectors $\mathbf{q}_2$:

$$V_{\mathbf{q}_1\mathbf{q}_2\mathbf{q}_3}^{TLL} = 2\frac{q_1 q_2^3}{q_3}\sin\Theta_2 \cos\varphi_2 \left(\frac{q_1}{2q_2} + \cos\Theta_2\right) \cdot \left\{ A_{cub}\left(1 + \frac{q_1}{q_2}\cos\Theta_2\right) + \right.$$
$$\left. + 2\widetilde{c}_{155}(\sin\Theta_2)^2(\cos\varphi_2)^2 + (2\widetilde{c}_{155} + \Delta C)\cos\Theta_2\left(\frac{q_1}{q_2} + \cos\Theta_2\right)\right\}, \tag{13}$$

$$\cos\Theta_2 = S* - \frac{q_1}{2q_2}(1 - S*^2), \quad S* = \frac{s_T}{s_L},$$

where $s_L$ and $s_T$ are the sound velocities for the longitudinal and transverse phonons, respectively. Within the long-wave approximation the relation (13) coincides with the Simpson result [25]. In order to verify in this coincidence we shall express the square of the matrix element averaged over the angle $\varphi_2$ in terms of the function $I_1(\theta)$, which is introduced in Ref. [25]:

$$\int_0^{2\pi} d\varphi_2 \left|V_{\mathbf{q}_1\mathbf{q}_2\mathbf{q}_3}^{TLL}\right|^2 = \left(\frac{q_1 q_2^3}{q_3}\right)^2 I_1(\theta),$$

$$I_1(\theta) = 4\pi(\sin\Theta_2 \cos\Theta_2)^2 \left(P_1 - M_1(\sin\Theta_2 \cos\Theta_2)^2 - N_1(\sin\Theta_2)^2\right), \tag{13a}$$

where the expressions for coefficients $P_1$, $M_1$, $N_1$ from Ref [25] have simpler from with the preceding notation:

$$P_1 = (c_{11} + c_{44} + 2c_{155})^2, \quad M_1 = 0.5 \cdot \left((2\widetilde{c}_{155} + \Delta C)^2 + (\Delta C)^2\right), \quad N_1 = P_1 \cdot (2\widetilde{c}_{155} + \Delta C) - M_1.$$

It is apparent that the coefficients $M_1$ and $N_1$ vanish at the passage to the isotropic medium model owing to condition (7). In expression (9), the integrals with respect to the angular variables $\Theta_2$ and $\varphi_2$ can be readily calculated. Indeed, the integral over $\Theta_2$ is calculated using $\delta$-function and taking into account the law of energy conservation for the three-phonon scattering (13a), while the integrals over $\varphi_2$ have nonzero contributions only due to the terms containing even powers of $\cos(\varphi_2)$). Eventually, we obtain the following expression the relaxation rate according to the Landau–Rumer mechanism:

$$v_{phN}^{TLL} = T^5 z_1 B^T, \quad B^T = \frac{(k_B)^5 S*^2(1 - S*^2)}{16\pi\hbar^4 \rho^3 s_T s_L^8} \frac{sh(z_1/2)}{(z_1/2)} J_z. \tag{14}$$

Here $J_z(z_1,T) = \int_{z_{min}}^{z_{dL}} dz F(z,z_1)\varphi_1(z,z_1)\{b_{iso} + b_{int} + b_{aniso}\}$,

$$F(z,z_1) = \frac{z^4\left(1+\frac{z_1}{2z}\right)^2}{\left(1+\frac{z_1}{z}\right)^2 sh\left(\frac{z}{2}\right) sh\left(\frac{z_1+z}{2}\right)}, \quad \varphi_1(z,z_1) = \left[1+\frac{z_1}{2z}\left(1+\frac{1}{S*}\right)\right]\left[1+\frac{z_1}{2z}\left(1-\frac{1}{S*}\right)\right],$$

$b_{iso} = (A\varphi_2(z,z_1))^2, \quad b_{int} = A\varphi_2(z,z_1)[2(2\tilde{c}_{155} + \Delta C)S*^2\varphi_3 + 3\tilde{c}_{155}(1-S*^2)\varphi_1(z,z_1)],$

$b_{aniso} = [(2\tilde{c}_{155} + \Delta C)S*^2\varphi_3]^2 + 3\tilde{c}_{155}(1-S*^2)\varphi_1(z,z_1)(2\tilde{c}_{155}+\Delta C)S*^2\varphi_3 + \frac{5}{2}[\tilde{c}_{155}(1-S*^2)\varphi_1]^2,$

$$\varphi_2(z,z_1) = 1 + \frac{z_1}{z} - \frac{1}{2}\left(\frac{z_1}{z}\right)^2\left(\frac{1}{S*^2}-1\right), \quad \varphi_3(z,z_1) = \left[1 - \frac{1}{2}\frac{z_1}{z}\left(\frac{1}{S*^2}-1\right)\right]\left[1+\frac{z_1}{2z}\left(1+\frac{1}{S*^2}\right)\right], \quad (14a)$$

where $z_1 = \frac{\hbar\omega_{qT}}{k_B T}$, $z_{min} = \frac{1}{2}(1/S^* - 1)z_1$, $z_{dL} = \frac{\hbar\omega_{dL}}{k_B T}$, $\omega_{dL}$ is the Debye frequency for the longitudinal phonons. The functions $\varphi_n$ ($n=1, 2, 3$) are related to with the angular dependence of the matrix element (13), and, at $z_1 \ll z$ they approach to 1, being zero at the values of $z$, equal to:

$$z_{10} = 0.5(1/S^* - 1)z_1, \quad z_{20} = \left((2/(S^*)^2 - 1)^{1/2} - 1\right)\cdot z_1, \quad z_{30} = 0.5\left[1/(S^*)^2 - 1\right]\cdot z_1. \quad (15)$$

This feature leads to the non-monotonic dependence of the phonon scattering probability on the ratio of the reduced wave vectors $z$ and $z_1$ at $z_1 > z$ (see section 4). In expression for the phonon relaxation rate (14) three contributions are included: the isotropic scattering $b_{iso}$; the interference of isotropic and anisotropic scattering $b_{int}$ and the contribution of anisotropic scattering $b_{aniso}$. In the isotropic case $\tilde{c}_{155} = \Delta C = 0$, therefore $b_{int} = b_{aniso} = 0$, and coefficient $b_{iso}$ becomes the form

$$b_{iso} = [A_{iso}\varphi_2(z,z_1)]^2. \quad (16)$$

For the cubic symmetry crystals in the long-wave limit $z_1/z \ll 1$ the formula (8) follows from the expressions (14) and (14a) with the coefficient $B_{T0}$ which for the temperatures significantly below the Debay temperatures is given by the expression

$$B_{T0} = \frac{B \cdot \pi^3 k_B^5 S*^2 (1-S*^2)}{15\hbar^4 \rho^3 s_T s_L^8}, \quad (17)$$

$$B = A^2 + A\cdot\{2(S*)^2(2\cdot\tilde{c}_{155} + \Delta C) + 3\cdot\tilde{c}_{155}(1-S*^2)\} + (2\tilde{c}_{155}+\Delta C)^2 S*^4 + \\ + 3\cdot(2\tilde{c}_{155}+\Delta C)\cdot\tilde{c}_{155}S*^2(1-S*^2) + \frac{5}{2}\cdot(\tilde{c}_{155})^2(1-S*^2)^2. \quad (17a)$$

Thus, in the long-wavelength limit the Landau–Rumer result for the transverse phonon relaxation rate is retained: $\nu_{phN}^{TLL} \approx z_1 T^5$. However, in addition to the isotropic scattering (the first term in expression (17a)) this mechanism also contains contributions due to the interference of the isotropic and anisotropic scattering (the second term in expression (17a)) and the anisotropic scattering (the last three terms in expression (17a)).

The estimates show, that contribution of the isotropic scattering in the [100] direction amount to approximately 0.1 % for Ge, 6 % for silicon and 52 % for diamond, the interference term gives 5 % for Ge and 40 % for diamond, and the contribution of anisotropic scattering amount to approximately 95 % for Ge and 8 % for diamond. In silicon the contribution of the interference scattering is negative, and its absolute value is a three times less than one for the anisotropic scattering. It is necessary to note, that for Ge, GaAs and diamond crystals the interference term is positive, since values $A_{cub}$ and $\tilde{c}_{155}$ are identical in signs (see. Table II). In contrast to Ge, GaAs and diamond, for Si, InSb, GaSb, CaF$_2$, SrF$_2$ crystals $A$ and $\tilde{c}_{155}$ have opposite signs, and the interference scattering for them gives the negative contribution to transverse phonon relaxation rate.

The values of parameters $B_{iso}$, $B$ и $B_{T0}$, and also coefficients of ultrasonic attenuation $\alpha^*_{TLL}$ for the crystals under consideration are presented in the Table III. The coefficient $\alpha^*_{TLL}$ normalized on factors $B_{T0}$ and $T^5$ is given by the expression:

$$\alpha^*_{TLL}(z_1,T) = \frac{\alpha_{TLL}(z_1,T)}{B_{T0}T^5}, \quad \alpha_{TLL}(z_1,T) = \nu^{TLL}_{phN}(z_1,T)/2s_T . \tag{18}$$

As seen from the comparison of parameters $B_{iso}$ and $B$, the greatest deviation from the model of isotropic medium takes place for crystals InSb, CaF$_2$, LiF for the diamond, GaAs and SrF$_2$ crystals these parameters differ by the value of 15-20 %. From Table III follows, that in the [100] direction at $T = 10$ K and $z = 1$ the $\nu^{TLL}_{phN}$ values are $1.2 \cdot 10^5$ s$^{-1}$ for Ge, $1 \cdot 10^4$ s$^{-1}$ for Si and $6.3 \cdot 10^3$ s$^{-1}$ for diamond.

**Table III.** Parameters determining the transverse phonon relaxation rates

|  | $B_{iso}$, $10^{24}$ dyn$^2$/cm$^4$ | $B$, $10^{24}$ dyn$^2$/cm$^4$ | | $B_{T0}$, s$^{-1}$K$^{-5}$ | | $\alpha^*$ | |
| --- | --- | --- | --- | --- | --- | --- | --- |
|  |  | [100] | [111] | [100] | [111] | [100] | [111] |
| Ge | 17,2 | 10,6 | 8,93 | 1,2 | 0,37 | 0,17 | 0,061 |
| Si | 16,27 | 8,9 | 7,54 | 0,099 | 0,035 | 0,0085 | 0,0034 |
| Diamond | 1737,7 | 1483,4 | 1443,1 | 0,0063 | 0,004 | 0,00025 | 0,0002 |
| InSb | 2,12 | 0,744 | 0,583 | 1,91 | 0,465 | 0,416 | 0,12 |
| GaSb | 10,2 | 5,38 | 4,34 | 3,73 | 1,01 | 0,67 | 0,218 |
| GaAs | 26,23 | 20,35 | 20,87 | 0,00013 | 0,00021 | 0,0007 | 0,0096 |
| CaF$_2$ | 4,26 | 8,34 | 7,65 | 0,15 | 0,37 | 0,023 | 0,046 |
| LiF | 0,71 | 2,18 | 2,82 | 0,139 | 0,061 | 0,014 | 0,0073 |
| SrF$_2$ | 3,47 | 4,44 | 4,37 | 0,207 | 0,33 | 0,029 | 0,042 |

The values of the fitting parameters $B_{T0}$, which were used in the analysis of the thermal conductivity for Ge and Si crystals with various isotope compositions in [4-6] (see table II in Ref. [10]), appear to be 43 times smaller for Ge and 38 times smaller for Si, than the corresponding results of the elasticity theory for coefficients $B_{T0}$. Since the relaxation rate $\nu^{TLL}_{phN}$ for thermal phonons ($z_1 > 1$) is significantly lower than that according to the Landau–Rumer theory in the entire temperature interval 1 K < $T$ < 100 K, the effective $B_{T0}$ value for thermal phonons has to be also smaller than that according to the long-wavelength approximation (see the next Section). However, even with allowance for these factors, the effective relaxation frequencies $\nu_{phN}^{TLL}$ obtained using the fitting parameters $B_{T0}$ for germanium and silicon crystals with different isotope compositions [5,9-10] are lower by one order of magnitude than the values obtained in our calculations within the framework of the theory of elasticity. This discrepancy between the results, received in the framework of the theory of elasticity and the results of the thorough fitting of the experimental data of the lattice thermal conductivity of Ge crystals with various isotope compositions [4-6,9,10] requires further analysis. The reason of this discrepancy could be clarify the experimental studies of the coefficient of the ultrasound absorption in structurally perfect and highly isotopically enriched Ge and Si crystals. It can be noted that the anisotropy of the relaxation rates $\nu_{phN}^{TLL}$ in the Ge and Si crystals are mostly due to the anisotropy of the second order elastic modulus.

In the isotropic case, formula (14) yields the result of Tucker and Rampton [14]: $B_{iso}=(A_{iso})^2$, where $A_{iso}$ is determined by the formula (12b). This expression is at variance with the result obtained for the isotropic medium by Maris (see Eq. (210) [24]):

$$B_{T0} = \frac{\pi^3 k_B^5}{120\hbar^4 s_L^5} \left( \frac{c_{11} - c_{44}}{c_{11}} \right) \cdot \left( \frac{c_{111} - c_{112} + 3c_{11} - c_{12}}{c_{11}} \right)^2. \tag{19}$$

Difference in numerical coefficients in expressions (19) and (17) by a factor of 2 is related to the two relaxation processes of traverses phonons taken into consideration in derivation of the formula (8). However, the main difference between expressions (14) - (18) and formula (19) consists in the elastic modulus $c_{111}$ from the last formula. The point is that this elastic modulus $c_{111}$ accounts for the relaxation of longitudinal phonons and is not related to the transverse phonon relaxation according to the Landau–Rumer mechanism. Evidently, the introduction of the generalized Gruneisen parameter [24] is inadequate procedure.

**4. The results of numerical analysis of the relaxation rate**

Let us analyze the dependence of the relaxation rate $\nu_{phN}^{TLL}(z_1,T)$ on the reduced wave vector $z_1$ and the temperature $T$ according to formulas (14) and (14a). First, we will consider the deviation from the Landau–Rumer formula (8) for the isotropic medium. The analysis of the ratio of rates $\nu_{phN}^{TLL}(z_1,T)$ and $\nu_{phN}^{LR}(z_1,T)$, designed according to formulas (14) and (16), has shown, that at the small $z_1$ this relation approach to unit and, accordingly, $\nu_{phN}^{TLL}(z_1,T) \to \nu_{phN}^{LR}(z_1,T)$. However as $z_1$ increases, the deviation of $\nu_{phN}^{TLL}(z_1,T)$ from $\nu_{phN}^{LR}(z_1,T)$ quickly increases, and at $z_1 \gg 1$ we have $\nu_{phN}^{TLL}(z_1,T) \ll \nu_{phN}^{LR}(z_1,T)$ in the entire temperature interval 1K < $T$ < 100K. Deviation of the relaxation rate from LR linear dependence (8) is most clearly shown for the function $\nu_{phN}^{TLL}(z_1,T)$. Figure 1 shows plots of the quantities:

$$\nu_{TLL}^{*}(z_1,T) = \frac{\nu_{phN}^{TLL}(z_1,T)}{B_{T0}T^5} \quad \text{and} \quad \nu_{TLL}^{*LR}(z_1) = z_1 \quad (20)$$

at low temperatures. As can be seen, a linear LR approximation (8) for $\nu_{TLL}^{*LR}(z_1)$ is possible only for $z_1 \leq 1$. For $z_1 > 4$ the value of $\nu_{TLL}^{*}(z_1,T)$ first sharply decreases, then in the interval $10 \leq z_1 \leq 20$ exhibits a "plateau", and for $z_1 > 20$ it is the monotonically decreasing function of $z_1$. Such strong deviation from the classical LR dependence (8) for the isotropic medium appeared unexpected. The analysis which has been carried out in works [26,27], has shown, that at $z_1 \gg 1$ ($\hbar\omega_{qT} \gg k_B T$) the relaxation rate $\nu_{phN}^{TLL}(z_1,T)$ exponentially decreases with increasing $z_1$ according to the expression:

$$\nu_{phN}^{TLL}(z_1,T) \propto (z_1)^n \exp(-0.5 \cdot (1/S^* - 1)z_1), \quad (21)$$

where $n=3$ according to [26, 14], and $n=5$ according to [27]. Therefore, it was possible to expect only a single maximum in curve $\nu_{phN}^{TLL}(z_1,T)$ and monotone decreasing for $z_1 \gg 1$. The occurrence of new feature - "plateau" at $10 < z1 < 20$ on the relaxation rate dependence on the phonons wave vector demands more careful analysis.

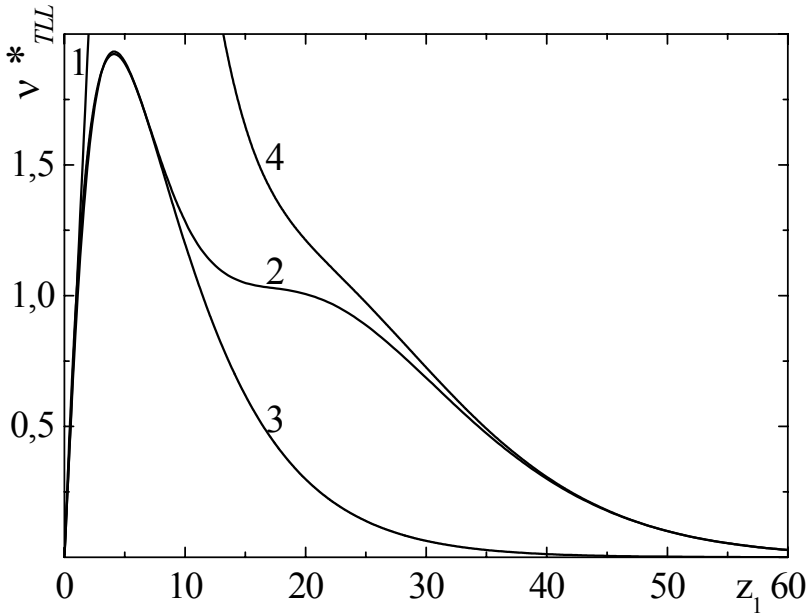

**Figure 1.** The dependences of $\nu_{TLL}^{*}(z_1,T)$ versus reduced wave vector $z_1$ at $T = 1$ K calculated using various models for the isotropic case: (*1*) linear Landau–Rumer approximation; (*2*) this study; (*3*) approximation (23); (*4*) approximation (22); all calculations performed for $s_L = 4.92 \cdot 10^5$ cm/s and $s_T = 3.55 \cdot 10^5$ cm/s.

Before proceeding with this analysis, it is necessary to make two important remarks. First, the observed peculiarities are related to the dependence of the matrix element (13) (the scattering probability) on the angle of phonon scattering. We strictly took this dependence into consideration, in contrast to Ref. [26-28]. Second, the characteristic shape of the $\nu_{TLL}^{*}(z_1,T)$ curve can be realized only for sufficiently low temperatures. Indeed, the limiting frequency of the transverse phonons is restricted to the Debye frequency, which amounts to approximately 118 K for germanium and 210 K for silicon. Therefore, although the characteristic shape of the dependence $\nu_{TLL}^{*}(z_1,T)$ is retained at high temperatures, the plateau in this curve at $T \sim 50\text{–}100$ K falls within a nonphysical region of the wave vectors exceeding the Debye wave vector: $z > z_{dT} = \hbar\omega_{dT}/k_BT$ ($\omega_{dT}$ is the Debye transverse phonon frequency). However, at low temperatures (about 1–4 K for germanium and 1–10 K for silicon), the aforementioned peculiarities fall within the physical region of the phonon wave vectors ($z \leq z_{dT}$).

Let's consider below the factors responsible for the appearance of the "plateau" on the plot of relaxation rate as a function of the phonon wave vector. The integrand function in Eq. (14) for the isotropic case comprises a product of three functions: $F^*(z,z_1) \cdot \varphi_1(z,z_1) \cdot (\varphi_2(z,z_1))^2$, where $F^*(z,z_1) = F(z,z_1) \cdot sh(z_1/2)$, and the functions $\varphi_1(z,z_1)$ and $\varphi_2(z,z_1)$ are related to the angular dependence of the scattering probability (see Eq. (13)). The dependences of these functions on $z$ at a fixed value of $z_1$ are shown in Fig. 2. The function $F^*(z,z_1)$ reaches a maximum at $z \sim 4$, and then exhibits exponential decay. The functions $\varphi_1(z,z_1)$ and $\varphi_2(z,z_1)$ tend to unity for $z_1 \ll z$ and are equal to zero when $z = z_{10}$ and $z = z_{20}$, accordingly.

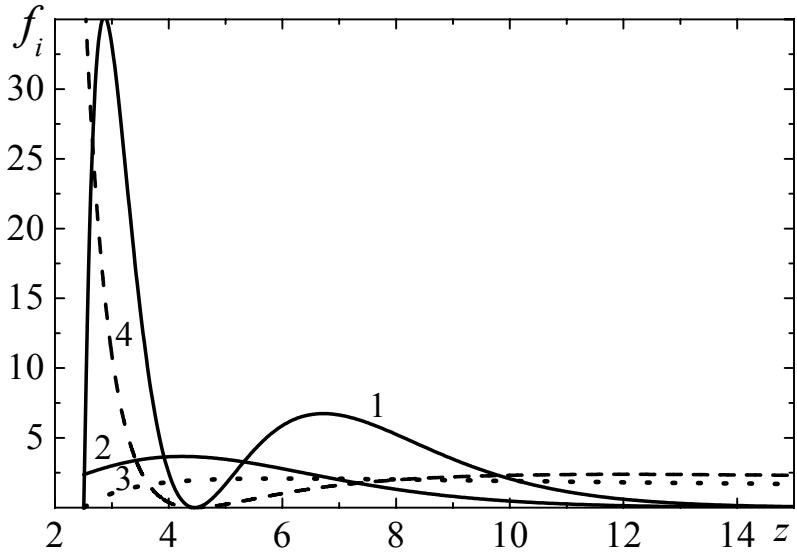

**Figure 2.** The dependences of the integrand functions versus reduced wave vector $z_1$: (*1*) $F^*(z, z_1)\varphi_1(z, z_1)\varphi_2^2(z, z_1)$; (*2*) $F^*(z, z_1)$; (*3*) $\varphi_1(z, z_1)$; (*4*) $\varphi_2^2(z, z_1)$. All calculations were performed for $z_1 = 13$ and the sound velocities $s_L = 4.92\ 10^5$ cm/s and $s_T = 3.55\ 10^5$ cm/s.

Evidently, the probabilities of phonon scattering by the angles $\Theta_1$ and $\Theta_2$ ($\cos(\Theta_1) = -1$, $\sin(\Theta_1)=0$ and $\cos(\Theta_2) = S^* - (1-S^*)(2S^*\alpha_{20})^{-1}$) are zero, which leads to a nonmonotone dependence of the scattering probability on the phonon wave vector (Fig. 2, curve 1). The function $(\varphi_2(z,z_1))^2$ quite rapidly increases on both sides of the point $z = z_{20}$. The value of $z_{20}$ increases with increasing of the parameter $z_1$ and, for $z = z_{20} = 4\text{–}5$, the function $\varphi_2(z,z_1)$ turns zero exactly in the region of maximum of the function $F^*(z,z_1)$. This leads to the appearance of inflection on the $\nu^*_{TLL}(z_1,T)$ curve and the onset of plateau at $z_1 \approx 16$ $z_{20} = 8\ [(2/(S^*)^2 - 1)^{1/2} - 1]^{-1}$. Further increase in $z_1$ leads to a shift of the zero of the function $\varphi_2(z,z_1)$ toward the region of exponential decay of function $F^*(z,z_1)$, whereby the values of $\nu^*_{TLL}(z_1,T)$ slightly decrease (which corresponds to the plateau) until the zero of the function $\varphi_1(z,z_1)$ would pass through the maximum of $F^*(z,z_1)$, which corresponds to the second inflection on the $\nu_{TLL}^*(z_1,T)$ curve (see Fig. 1). At $z_{min} > 4$ or $z_1 > 16\, z_{10} = 8\cdot(1/S^* - 1)^{-1}$, the transverse phonon relaxation rate occurs in the region of exponential decay. Thus, it is evident that the angular dependence of the phonon scattering probability should be taken into account in calculations of the phonon relaxation rates.

In the region of $z_1 \gg 1$ and $z_{min}, z_{20} \gg 1$, the value of the preexponential factor can be estimated as follows. The main contribution to the integral is related to the region of the first peak $z_{min} < z < z_{20}$ (see Fig. 2). According to the theorem of the mean, functions smoothly varying, can be put before the integral over this region at the middle point $\bar{z} = (z_{min} + z_{20})/2$, while the remaining strongly varying part $\varphi_1 \cdot \varphi_2^2 \exp(-z)$ has to be precisely integrated. This yields:

$$\nu_{phN}^{TLL}(z_1,T) = 7 \cdot \left(\frac{\bar{z}+0.5}{\bar{z}+1}\right)^2 \left(\bar{z}+\frac{1}{2}(\frac{2}{S^*}-1)^{1/2}+1\right)^2 (\bar{z}+\frac{1}{2}(\frac{1}{S^*}+1))\left\{(\alpha_{20}-\alpha_{10})^2 z_1^3 - 4(\alpha_{20}-\alpha_{10})z_1^2 + 6z_1\right\}\exp(-\alpha_{10}z_1). \quad (22)$$

As can be seen from Fig. 1 (curve *4*), this approach provides a reliable estimate of $\nu_{phN}^{TLL}(z_1,T)$ for the high-frequency phonons. Thus, allowance for the angular dependence of the phonon scattering probability (13) leads, in contrast to the results obtained in [26,27], to a preexponential factor in the form of the third-order polynomial, which cannot be reduced to a simple power dependence (21).

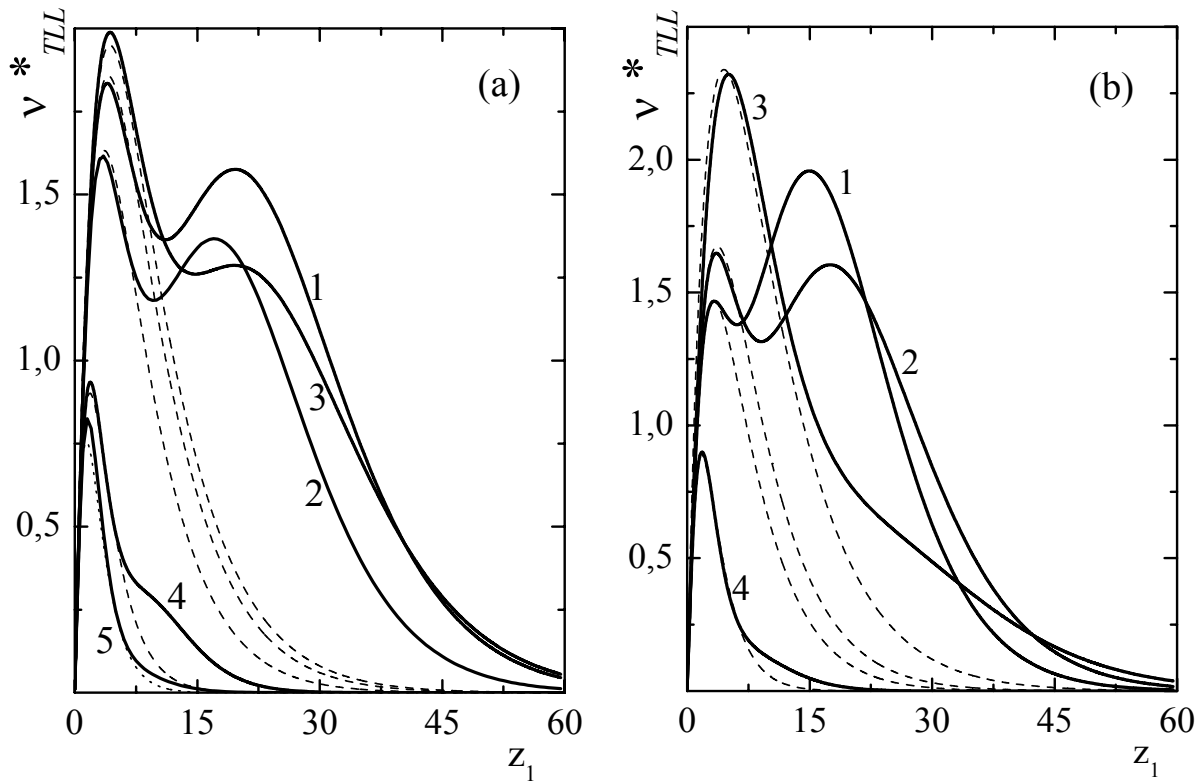

**Figure 3.** The dependences of the phonon relaxation rate versus reduced wave vector $z_1$ in the [100] direction for crystals: (a) *1* - Ge, *2* - Si, *3* - diamond, *4* - GaAs and *5* - CaF$_2$; (b) *1* - InSb, *2* - GaSb, *3* - LiF$_2$ and *4* - SrF$_2$ at $T = 1$ K. Broken curve shows the approximation by formula (23).

In contrast to the isotropic case, the angular dependence of the squared matrix element for the cubic crystals is determined by a combination of three functions: $\varphi_1(z,z_1)$, $\varphi_2(z,z_1)$ и $\varphi_3(z,z_1)$. In this case, a change in positions of the zeros of these functions ($z_{10} = z_{min}$, $z_{20}$ and $z_{30}$) relative to the maximum of $F^*(z,z_1)$ with increasing the phonon wave vector $z_1$ also determines peculiarities in the behavior of the relaxation rate $v_{TLL}^*(z_1,T)$. For the cubic crystals the probability of phonon scattering turns zero only at the lower integration limit $z_{10}= z_{min}$ ($\theta_2 = \pi$).

However, separate contributions to the scattering probability may also turn zero: the isotropic scattering contribution is zero for $z=z_{20}$; the interference contributions and some of the terms corresponding to the anisotropic scattering of phonons and proportional to the functions $\varphi_2(z,z_1)$ and $\varphi_3(z,z_1)$ are zero at the points $z=z_{20}$ и $z= z_{30}$.

Let us consider the dependence of the phonon relaxation rate on the reduced wave vector in the [100] direction. As can be seen from Figs. 3а and 3б, these dependences qualitatively differ from those obtained in the case of the isotropic medium: for Ge, Si, diamond, InSb и GaSb crystals they are nonmonotone and display a minimum and the second maximum on the $v_{TLL}^*(z_1,T)$ dependences instead of the plateau. However for GaAs, $CaF_2$, $LiF_2$ and $SrF_2$ crystals the second maximum is absent, and for the semiconductor GaAs crystal instead of "plateau" the interval of slower decrease of the relaxation rate takes place. The appearance of new peculiarities in the dependence of the phonon relaxation rate on the phonon wave vector - the minimum and the second maximum at $z_1 \gg 1$ for Ge, Si, diamond, InSb и GaSb crystals is of considerable interest from the standpoint of ultrasonic investigations and requires more thorough investigation.

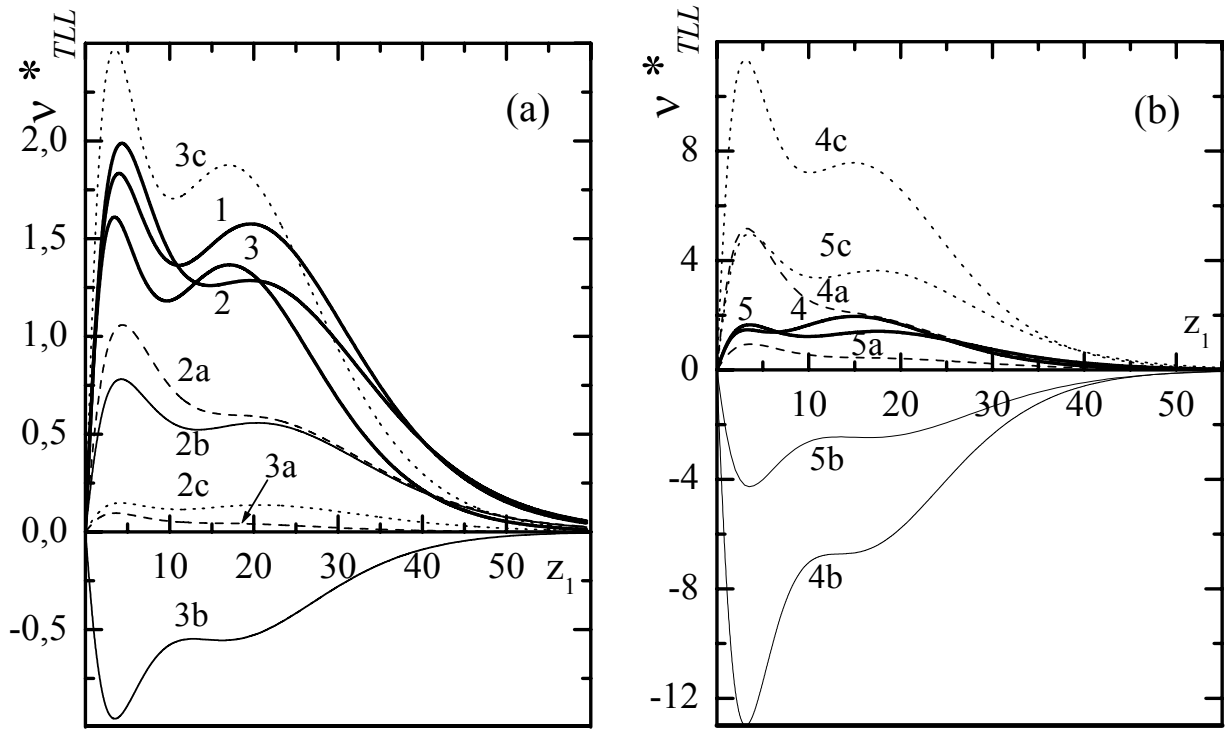

**Figure 4.** The dependences of the phonon relaxation rate $v_{TLL}^*$ in the [100] direction versus the reduced wave vector $z_1$ at low temperatures: curves 1, 2, 3, 4, and 5 for Ge, diamond, Si, InSb and GaSb, respectively. Curves 2a, 3a, 4a and 5a are the contributions of the isotropic scattering; curves 2b, 3b, 4b and 5b are the contributions of the interference of isotropic and anisotropic scattering; curves 2c, 3c, 4c and 5c - the contribution of anisotropic scattering.

The analysis of all contributions to the relaxation rate $\nu^*_{TLL}(z_1,T)$ has shown that the occurrence of the second maximum is related to the cubic anisotropy of crystals. The interference of isotropic and anisotropic scattering (the second term in expression (14a)), as well as the anisotropic scattering (the last three terms in expression (14a)) causes a two-peak character of the dependence of $\nu^*_{TLL}(z_1,T)$ on the reduced wave vector in Ge, Si, diamond, InSb and GaSb (see Figs. 4a and 4b, the curves 2b, 3b, 4b, 5b and 2c, 3c, 4c, 5c). All three contributions for Ge and diamond have a positive sign (see Fig. 4a, curves 2a, 2b and 2c), since for them $A$ and $\tilde{c}_{155}$ have the same sign (see Table II). Our analysis has shown that for Ge and Si the dominant contribution to the relaxation rate is due to the anisotropic scattering, whereas for diamond, in the vicinity of the first maximum, the dominant contribution is due to the isotropic scattering. For these crystals the contribution of the anisotropic scattering decreases from 95 % for Ge down to 8 % for diamond, and the contribution of the isotropic scattering increases from 0.1 % for Ge to approximately 40 % for diamond. For silicon the contribution of the interference scattering is negative, its absolute value being approximately three times smaller than that of the anisotropic scattering contribution (see Fig. 4a, curves 3b and 3c). Therefore the second maximum on the dependence of the relaxation rate is more pronounced for Ge and Si crystals being much weaker for diamond.

In contrast to Ge and diamond, for Si, InSb and GaSb crystals $A$ and $\tilde{c}_{155}$ have opposite sings, and the interference scattering for them gives the negative contribution to the transverse phonon relaxation rate. For InSb, in contrast to Si, the contribution of the interference scattering essentially exceeds the contribution of the anisotropic scattering, so that the total contribution of the isotropic and anisotropic scattering in a vicinity of the first maximum is essentially compensated by the interference scattering (see Fig. 4b, curves 5a, 5b and 5c). However, for GaSb the contribution of the interference scattering is slightly less than the contribution of the anisotropic scattering, and such a compensation of contributions in the vicinity of the first maximum does not occur (see Fig. 4b, curves 4a, 4b and 4c). Therefore for GaSb, in contrast to Ge, Si and diamond, the amplitudes of the first and the second maximums of the relaxation rate dependence $\nu^*_{TLL}(z_1,T)$ are close to each other, whereas for InSb crystals the amplitude of the second maximum of the phonon relaxation rate exceeds that of the first one by a factor 1.3 (see also Fig. 5a and 5b).

It should be noted that the dependence of the relaxation rate $\nu^*_{TLL}(z_1,T)$ on the phonon wave vector for Ge, Si, diamond and GaSb crystals in the [111] direction is similar to the one for the isotropic case (see fig.2): the second maximum is not observed, and the region of "plateau" exists for these crystals. In contrast to this, for crystals GaAs, $CaF_2$, LiF and $SrF_2$ instead of the region "plateau" the interval of slower decrease of the relaxation rate takes place. In contrast to these crystals, for InSb in the [111] direction curve of relaxation rate dependence has a two-peak character and the amplitudes of the first and second maximum being close to each other. Thus, the appearance of the second maximum on the $\nu^{*}_{TLL}(z_1, T)$ curves for the [100] direction in the crystals of germanium, silicon, diamond, InSb and GaSb is related to the cubic anisotropy of these crystals.

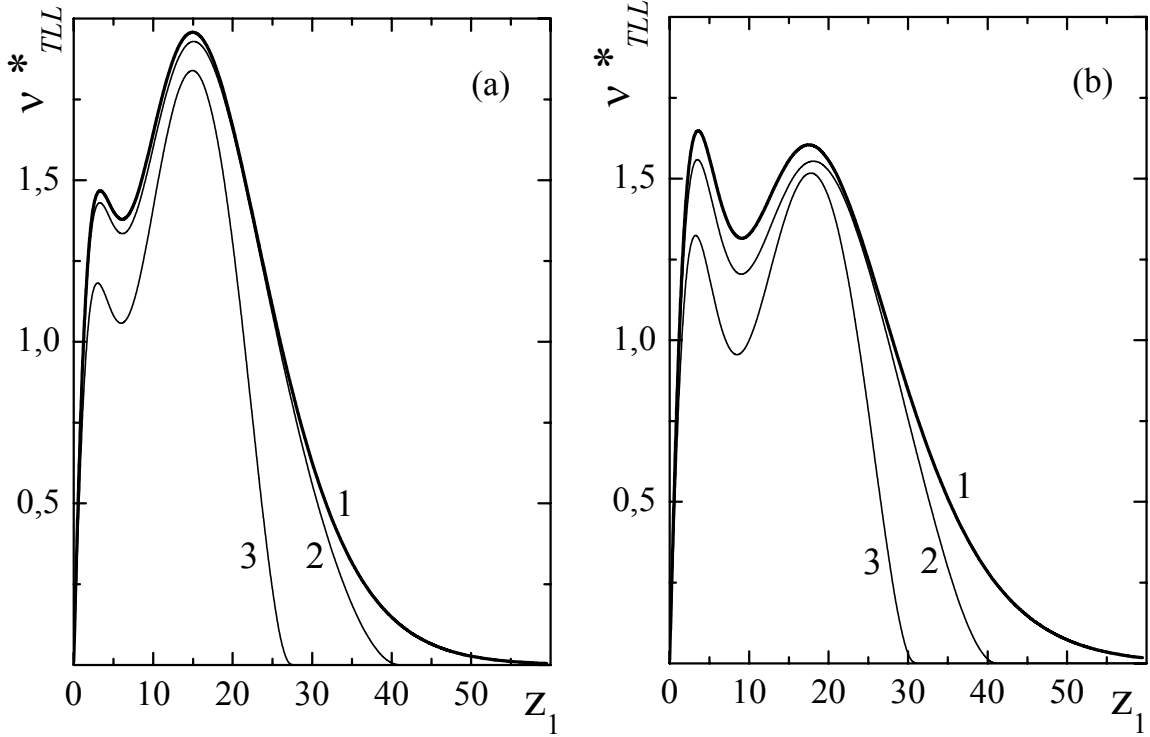

**Figure 5.** The dependences of the phonon relaxation rate versus reduced wave vector $z_1$ in the [100] direction at various temperatures for crystals: (a) InSb $T$ = 1–10 K (*1*), 20 K (*2*), and 30 K (*3*); (b) GaSb at $T$ = 1–20 K (*1*), 30 K (*2*), and 40 K (*3*)

As for the analysis of thermal conductivity of the considering crystals, the use of formulas (14) for $\nu^{TLL}_{phN}(z_1, T)$ is rather inconvenient and strongly complicates these calculations. Since the actual energies of photons important for of thermal conductivity are limited from above on a level of $z1 < 4$–$5$ (greater values are exponentially cut by the Planck distribution function), the phonon relaxation rate in the region of the first maximum can be approximated by the expression:

$$\nu^{TLL}_{phN}(z_1, T) \approx \beta \cdot B_{T0} T^5 z_1 [\exp(-\alpha_{10} z_1) + \exp(-\alpha_{20} z_1)], \tag{23}$$

where coefficient β = 0,65 for Ge, Si, InSb, GaSb, LiF and diamond crystals, for GaAs β = 0.71, for $SrF_2$ and $CaF_2$ β = 0.78. In the region of $0 < z1 < 8$ (see Fig. 3a and 3b), this approximation well agrees with the results obtained with formulas (14) for both the isotropic case and the cubic crystals. At least, this formula provides a better approximation than the Landau–Rumer formula (8) with a fitting parameter.

Let us also consider the possibility of experimental observation of the aforementioned features in the dependence of the transverse phonon relaxation rate on the wave vector. At the low temperatures the value of relaxation rate $\nu^{*}_{TLL}(z_1, T)$ for Ge, Si, InSb, GaSb and diamond crystals is practically a function of only the variable $z_1$. As can be seen from Fig. 5, for crystals InSb and GaSb it very slightly changes depending on the temperature. The positions of maxima and minima, as well as their absolute values, vary within less than 1% in the temperature interval 1 K < $T$ < 20 K for Ge, InSb and GaSb; 1 K < $T$ <

60 K for silicon, and 1 K < $T$ < 100 K for diamond. These features allow the dependence of value $\nu^*_{TLL}(z_1, T)$ on a phonons wave vector to be determined from measurements of the ultrasound absorption coefficient. The coefficient of ultrasound absorption $\alpha_{TLL}$ is proportional to the phonon relaxation rate $\nu^{TLL}_{phN}$ (see (18)). Therefore, if the absorption coefficient of ultrasound with the energy of $\hbar\omega_{dT} \approx 10\,K$ is measured in the temperature interval from 0.1 to 50 K, it is possible to determine the behavior of $\nu^*_{TLL}(z_1, T)$ in the interval of wave vectors $0.1 < z1 < 100$ as:

$$\nu^*_{TLL}(z_1, T) = \frac{\alpha_{TLL}(z_1, T) 2 s_T}{B_{T0} T^5}. \tag{24}$$

For this purpose, it is necessary to transform the temperature dependence $\nu^*_{TLL}(z_1, T)$ at a fixed phonon frequency $\omega_{qT}$ into the function $z_{qT} = \hbar\omega_{qT}/k_B T$ at a fixed temperature, taking into account that, in the interval of low temperatures indicated above, the rate $\nu^*_{TLL}(z_1, T)$ is practically a function of only $z_1$. In silicon and diamond crystals, the observations conditions are more favorable for transverse phonons in the terahertz range (1 THz ≈ 50 K). Thus, the dependence of the relaxation rate on the phonon frequency in cubic crystals can be determined from the results of measurements of the ultrasound absorption coefficient. It can be recommended to perform such experiments with highly isotopically enriched, pure crystals of germanium, silicon, and diamond in order to reduce the phonon scattering on defects, which can mask the anharmonic phonon scattering processes at rather low temperatures.

## 5. Conclusion

Thus, we have considered the relaxation of transverse thermal and high-frequency phonons for the Landau–Rumer mechanism in the isotropic medium and in crystals of germanium, silicon, diamond, InSb, GaSb, GaAs, CaF$_2$, LiF and SrF$_2$ possessing a cubic symmetry. The energy of the elastic deformation caused by anharmonicity of vibrations of a cubic crystal lattice is expressed in terms of the second- and third-order elastic moduli. Using the known values of these elastic moduli, the parameters determining the transverse phonon relaxation rates for the Landau–Rumer mechanism have been evaluated for the crystals under consideration. It is shown that the dependence of the relaxation rate on the wave vector of thermal and high-frequency phonons strongly differs from the classical Landau–Rumer relationship both in the isotropic medium and in the cubic crystals studied. In contrast to the isotropic medium, for Ge, Si, InSb, and GaSb crystals in the [100] direction, these dependences have essentially non-monotonic character with two maxima, the second of which is in the range of high-frequency phonons $\hbar\omega_{qT} \gg k_B T$.

This anomaly is most pronounced for InSb crystals. In contrast to Ge, Si, diamond and GaSb, for InSb in the [100] direction the amplitude of the second peak on the wave vector dependence of the phonon relaxation rate essentially exceeds that of the first one, and in the [111] direction this dependence has two peaks with nearly equal amplitudes. For other studied crystals, in the [111] direction the phonon relaxation rate dependences are similar to the isotropic case: instead of the second peak, there is a region of slower decrease of the relaxation rate. The features of the relaxation rate are found to result from the angular dependence of the probability of anharmonic scattering and from the anisotropy of elastic properties. We have determined the values of scattering angles for which the scattering probability vanishes. We have also considered the possibility of experimental determination of the relaxation rate of high-frequency phonons as a function of the phonon wave vector from measurements of the temperature dependence of the absorption coefficient of high-frequency ultrasound.

We are grateful to I.F.Mirsaev, V.I.Ozhogin, A.V.Injushkin and A.N.Taldenkov for discussion of our results and for critical comments.

**Acknowledgment**

This work is supported by the grants from the President's of the RF, no. S.S. 1380.2003.2, no Y.D.-4130.2004.2, from the Dinastiya and from MTsFFM Foundations and the Russian Foundation for Basic Research "Ural" no.15-02-04.